\documentclass{aa}
\usepackage{psfig}

\def\bron{SAX J2239.3+6116}
\def\egret{3EG~J2227+6122}
\def\ecs{erg~cm$^{-2}$~s$^{-1}$}

\def\lum{erg~s$^{-1}$}

\begin{document}
\thesaurus{05(08.09.2 \bron, 4U~2238+60, M~2233+595, 3A~2237+608,
AT~2238+584, 3EG~J2227+6122)}

\title{The transient X-ray source \bron\ and its optical counterpart}
\titlerunning{The transient \bron\ and its optical counterpart}
\author{J.J.M.~in~'t~Zand\inst{1}
\and J.~Halpern\inst{2}
\and M.~Eracleous\inst{3,4}
\and M. McCollough\inst{5}
\and T. Augusteijn\inst{6}
\and R.A. Remillard\inst{7}
\and J. Heise\inst{1}
}
\offprints{J.J.M.~in~'t Zand (at e-mail {\tt jeanz@sron.nl})}

\institute{     Space Research Organization Netherlands, Sorbonnelaan 2,
                NL - 3584 CA Utrecht, the Netherlands
     \and
                Columbia Astrophysics Laboratory, Columbia University,
                New York, NY 10027, U.S.A.
     \and
                Department of Astronomy and Astrophysics , Pensylvania
                State University, 525 Davey Lab, University Park, PA 16802,
                U.S.A.
     \and       Visiting Astronomer, Kitt Peak National
                Observatory, National Optical Astronomy Observatories, which
                is operated by AURA, Inc., under a cooperative agreement with
                the National Science Foundation.
     \and       Universities Space Research Association, Huntsville, AL 35805, 
                U.S.A.
     \and       Isaac Newton Group of Telescopes, Apartado 321, 
                38700 Santa Cruz de La Palma, Canary Islands, Spain
     \and       Center for Space Research, Massachusetts Institute of
                Technology, Room 37-595, Cambridge, MA 02139, U.S.A.
                        }
\date{Received, accepted }
\maketitle

\begin{abstract}
We have discovered with the Wide Field Cameras on board {\em BeppoSAX\/}
the weak transient X-ray source \bron\ whose position coincides with that of
4U~2238+60/3A~2237+608 and is close to that of the fast transient AT~2238+60
and the unidentified EGRET source 3EG~2227+6122. The data suggest that the
source exhibits outbursts that last for a few weeks and peak to a
flux of $4\times10^{-10}$~\ecs\ (2-10~keV) at maximum. During the peak the
X-ray spectrum is hard with a photon index of $-1.1\pm0.1$. Follow-up
observations with the Narrow-Field Instruments on the same platform revealed
a quiescent emission level that is 10$^3$ times less. Searches through
the data archive of the All-Sky Monitor on RXTE result in the recognition
of five outbursts in total from this source during 1996-1999,
with a regular interval time of 262~days.
Optical observations with the KPNO 2.1~m telescope provide
a likely optical counterpart. It is a B0~V to B2~III star with
broadened emission lines at an approximate distance of 4.4~kpc.
The distance implies a 2-10~keV luminosity in the range from $10^{33}$ to
$10^{36}$~\lum. The evidence
suggests that \bron\ is a Be X-ray binary with an orbital period of
262 days.
\end{abstract}

\section{Introduction}
\label{intro}

The Wide Field Cameras (WFCs) on board the {\em BeppoSAX\/} satellite
carry out an extensive monitoring program of the whole X-ray sky.
Various kinds of transient X-ray sources are detected during these
observations on time scales between a fraction of a second and months. We here
discuss one case which we have followed up with the more sensitive 
Narrow-Field Instruments (NFI) on board {\em BeppoSAX} and with
the Kitt Peak National Observatory (KPNO) 2.1~m optical telescope,
and whose signature we searched for in data from the All-Sky Monitor (ASM)
on the Rossi X-ray Timing Explorer (RXTE) and in occultation data from
the Burst And Transient Source Experiment (BATSE) on the Compton
Gamma-Ray Observatory. In all observations we have interesting detections.
We describe analyses of WFC (Sect. \ref{secwfc}), NFI (Sect.~\ref{secnfi}),
ASM (Sect.~\ref{secasm}), BATSE (Sect.~\ref{secbatse}) and optical data
(Sect.~\ref{secopt}), point to possible associations
of previously found sources of high-energy emission in the field
(Sect.~\ref{secoth}), and discuss the possible nature of the source 
(Sect.~\ref{secdiscuss}).

\section{WFC observations}
\label{secwfc}

The WFCs (Jager et al. 1997) on the {\em BeppoSAX\/} satellite (Boella et al.
1997a) carry out a monitoring program where they point at arbitrary positions
in the sky. This program is the 'secondary mode' program whereby the
WFC pointings are dictated by those of the Narrow Field Instruments (NFI)
on the same satellite and the constraints of the satellite attitude with
respect to the direction of the sun. The secondary mode observations 
entail roughly 90\% of all observations. 

There are two identical WFCs. They each have $40\times40$~square degrees
field of views (full width to zero response) with $\sim5$\arcmin\ angular
resolution. The bandpass is 2 to 26 keV. The WFCs have been active
continuously since mid-1996, except for a 1-month period in May 1998 when 
unit 1 was turned off, and a 3 week period in early 1999 when both cameras
were turned off. During mid-May to mid-August 1997, the satellite was
in a standby mode, with both cameras turned on but without sufficient
stabilization for sensitive imaging.

During 103.4~d on-source time with a net exposure time of 27.8~d, 
the secondary observations
covered the field around the supernova remnant Cas~A. This exposure
is roughly uniformly distributed in time. The latest data that we
report here were obtained
on Dec. 15, 1999. Except for Cas~A,
there are no bright X-ray sources present in this field. Compared to the
background levels in the WFCs, Cas~A is not so bright. The sensitivity is, 
therefore, close to optimum in the field. During two instances, a relatively 
faint transient turned up in the field, at an
angular distance of 6.0\degr\ from Cas~A. They were on March 4, 1997,
and May 8, 1999. We note that sometimes Cyg~X-2 is in the same field of
view as this transient, but at an angular distance of 24.5\degr\ which
implies it does not degrade the signal of the transient.
The celestial position of the WFC-detected transient is
$\alpha_{\rm 2000.0}=~22^{\rm h}$39$^{\rm m}$22.8$^{\rm s}$, 
$\delta_{2000.0}$~=~+61\degr16\arcmin47\arcsec\
with a 99\% confidence
error radius of 1\farcm8.

\begin{figure}[t]
\psfig{figure=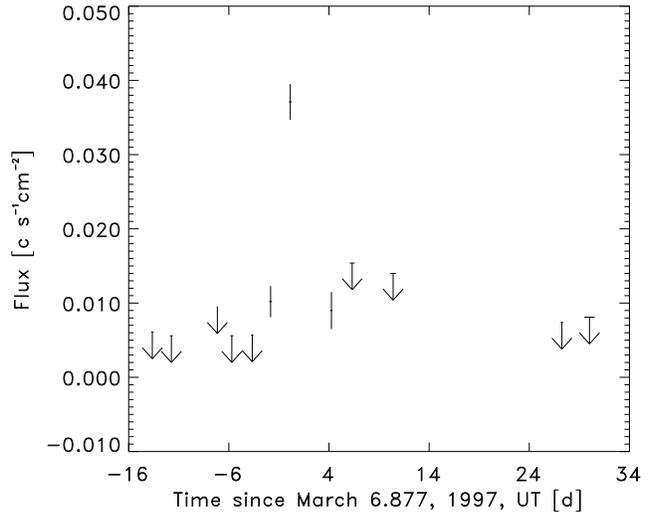,width=\columnwidth,clip=t}

\caption[]{Light curve of the first WFC-detected outburst of \bron\ in
March, 1997, 
for 2 to 26 keV with a time resolution set by the duration of the measurements
(these are of order 1~d).
The arrows indicate $3\sigma$ upper limits. These limits vary considerably
because of varying exposure times and varying exposed detector areas.
March 6.877, 1997, is MJD~50513.877.
\label{figwfclcop}}
\end{figure}

The detection in March, 1997, was the strongest WFC detection obtained
so far. In Fig.~\ref{figwfclcop}, we present a light curve with a resolution
of about a day. The period
over which we detected activity is about 1 week. The variation in
brightness during this event
seems to suggest that it was not active for much longer.

\begin{figure}[t]
\psfig{figure=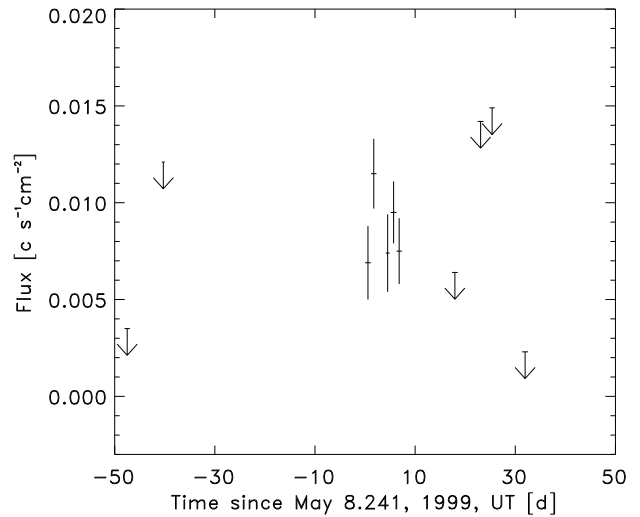,width=\columnwidth,clip=t}

\caption[]{Light curve of the second WFC-detected outburst of \bron\
for 2 to 26 keV. May 8.241, 1999, is MJD~51306.241.
\label{figwfclcop2}}
\end{figure}

A light curve of the second WFC-detected 
outburst is presented in Fig.~\ref{figwfclcop2}.
The measured peak flux of this outburst is about a factor of 3 lower
than that of the first outburst but this could very well be a sampling effect.
The source was above the detection threshold for one week but due to
an unfavorable sampling it could just as well have been active for 50 days.

\begin{figure}[t]
\psfig{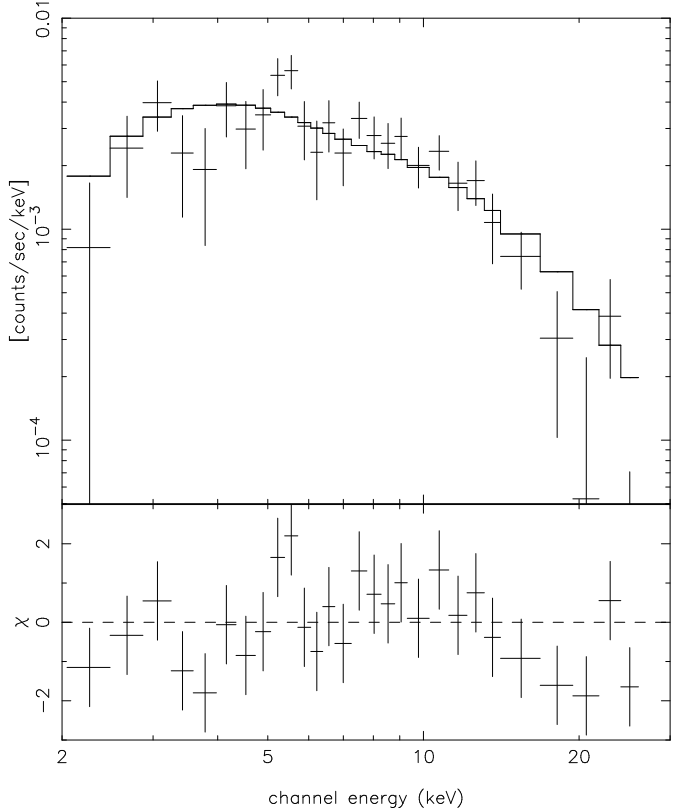}

\caption[]{Top panel: WFC-measured spectrum of \bron\ (crosses) and the 
fitted power law model (histogram). Bottom panel: residuals between
data and model in units of $\sigma$.
\label{figwfcspectrum}}
\end{figure}

We have extracted the spectrum from the observation at the peak of the first
outburst and modeled it
with a simple power law plus absorption due to interstellar gas of cosmic
abundances (according to Morrison \& McCammon 1983). We kept the value
for the hydrogen column density $N_{\rm H}$ fixed at 
$1.0\times10^{22}$~cm$^{-2}$, as found from HI maps (Dickey \& Lockman 1990).
The fit was satisfactory, with $\chi^2_{\rm r}=1.22$ (26 dof). The photon
index is quite hard at $-1.1\pm0.1$. The spectrum is shown in
Fig.~\ref{figwfcspectrum}. The 2-10 keV flux is $3.3\times10^{-10}$~\ecs\
which is 0.016 times the flux of the Crab in the same bandpass. In the
energy range 2 to
26 keV the flux is $1.0\times10^{-9}$~\ecs\ or 0.033 times that of the 
Crab.

\begin{figure}[t]
\psfig{figure=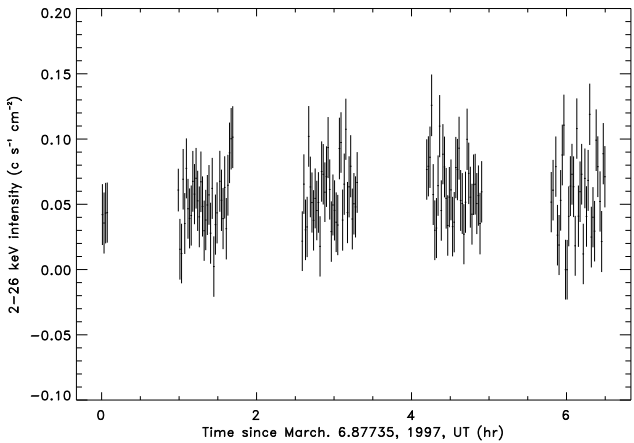,width=\columnwidth,clip=t}

\caption[]{2 to 26 keV light curve of \bron\ on March 6, 1997. The
time resolution is 75~s.
\label{figwfclc}}
\end{figure}

In Fig.~\ref{figwfclc}, we present a light curve with a relatively
high time resolution when the flux was at maximum during the first outburst.
The source appears to be variable but no coherent signal
could be found in this data set. The 90\%-confidence upper limit on the
amplitude of a periodic sinusoidal variation is 10\% for periods between
1 and 10$^3$~s.

We have analyzed WFC data taken on December 13-15, 1999. This is 2 weeks
after the optical measurements described in Sect. \ref{secopt}. 
No X-ray detection was found. The upper limit on the X-ray
flux is 6$\times10^{-11}$~\ecs\ (2-10 keV).

\section{NFI observations}
\label{secnfi}

\begin{figure}[t]
\psfig{figure=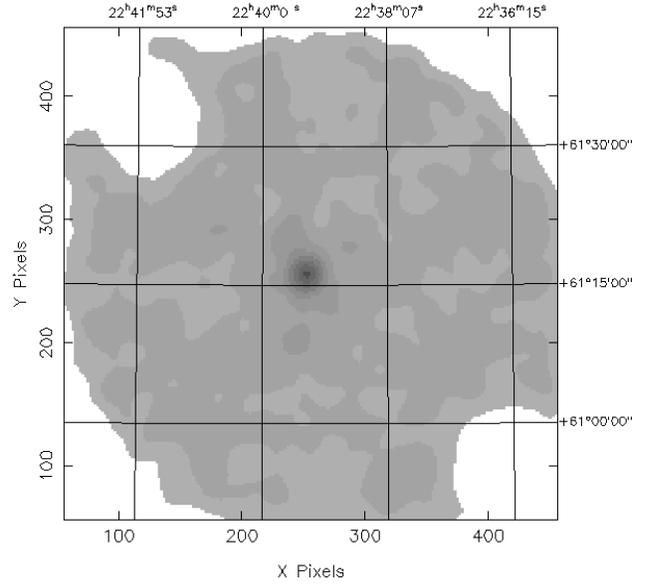,width=\columnwidth,clip=t}

\caption[]{MECS image of observation on \bron. MECS units 2 and 3
data are combined.
\label{fignfipicture}}
\end{figure}

Observations were made with the NFI on Nov. 
24.46-25.04 (UT), 1998, which is 627~d after the first and 165~d before 
the second WFC detection. The NFI include the
 low-energy concentrator/spectrometer (LECS; 0.1-10~keV; Parmar
et al. 1997), the medium-energy concentrator/spectrometer (MECS;
1.8-10~keV; Boella et al. 1997b),
the high-pressure gas scintillator proportional counter (HP-GSPC;
4-120~keV; Manzo et al. 1997), and the phoswich detector system
(PDS; 15-200~keV, Frontera et al. 1997). The first two instruments
are imaging devices. The effective exposure times
on \bron\ are 11.9~ks (LECS), 20.1~ks (MECS)
and 10.0~ks (PDS). The HP-GSPC was turned off for operational reasons.

Only the MECS data show a detection of \bron\ in an otherwise
empty field of view (see Fig.~\ref{fignfipicture}). The position of
the source is 
$\alpha_{2000.0}$~=~22$^{\rm h}$39$^{\rm m}$18.5$^{\rm s}$, 
$\delta_{2000.0}$~=~+61\degr16\arcmin20\arcsec\ with a 90\% confidence
error radius of 1\farcm0. This is 0\farcm7 from the WFC centroid
position and the two positions are fully consistent (see Fig.~\ref{figopmap}).
The chance probability of a background point source with a flux in
excess of $10^{-13}$~\ecs\ in the WFC error region is 0.3\% (based on the
counts by Giommi et al. 1999). We reduced the spectrum of the point source
from the MECS data through application of
the maximum likelihood method (e.g., In~'t~Zand et al. 2000)
and tried to model it with absorbed single-component models like a
power-law function and thermal bremsstrahlung, while fixing $N_{\rm H}$
to 1$\times10^{22}$~cm$^{-2}$. Only black-body radiation
gave a satisfactory fit to the data of a single-component model, with
$\chi^2_{\rm r}=1.74$ for 7 dof. The color temperature is $1.2\pm0.2$~keV
and the emission area at 4.4~kpc (see Sect.~\ref{secopt}) for isotropic
emission is equal that of a sphere with a radius of $0.6\pm0.4$~km.
The 2-10 keV flux is
$5\times10^{-13}$~\ecs\ which is $\sim10^3$ times fainter than
the peak flux of the March 1997 outburst.

\section{ASM observations}
\label{secasm}

\begin{figure}[t]
\psfig{figure=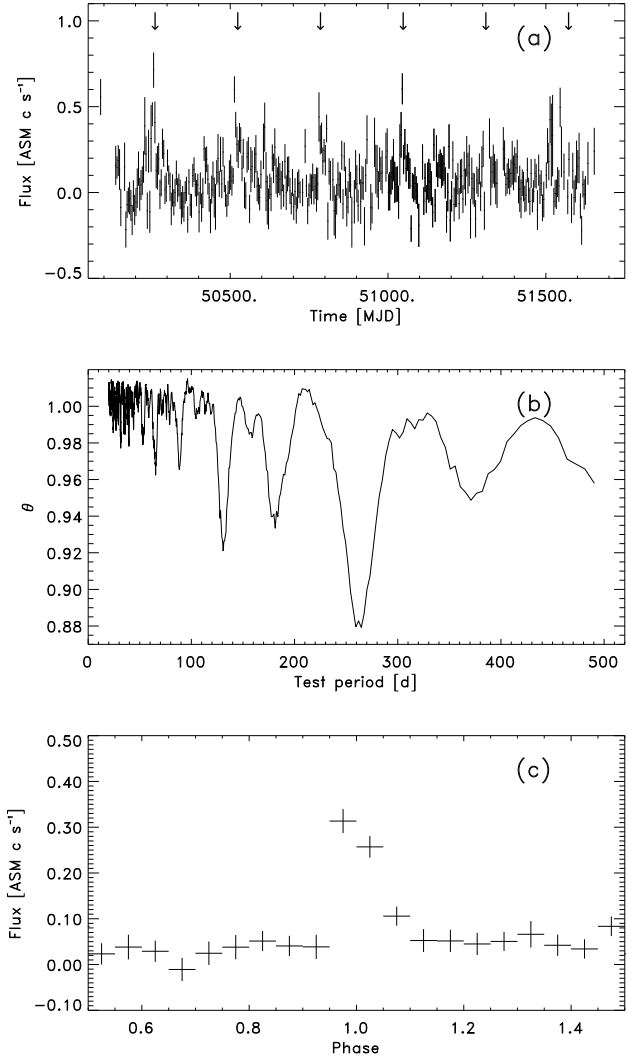,width=\columnwidth,clip=t}

\caption[]{(a) ASM light curve at 4~d time resolution, the arrows
point to the predicted times of outbursts for periodicity of 262 d;
(b) $\theta$ statistic (Stellingwerf 1978) as a function of test
period; (c) folded light curve at the best fit period. The non-zero
flux outside the outburst points to an instrumental bias level of
about 0.5~mCrab.
\label{figasm}}
\end{figure}

Given the NFI position, we have searched for signatures of \bron\ in
archival ASM data. The ASM is fully operational since March 1996 and
monitors each position on the sky in 2 to 12 keV during 90~sec snapshots
with a frequency of 5 to 10 times a day at a sensitivity of $\sim10$~mCrab
per day of observervations on uncrowded fields (Levine et
al. 1996). The lightcurve for \bron\ at 4-day resolution is presented
in Fig.~\ref{figasm}a. There is the suggestion for a detection during 
5 instances, at regular interval times of about 262~d. We tested
this periodicity by first filtering out the data within 10~d of closest
aproaches to the Sun and all data after MJD~51450 because there does
not appear to be an outburst there, and then calculating
the variance statistic
$\theta$ as defined by Stellingwerf (1978) for a range of test periods, see
Fig.~\ref{figasm}b. The resulting period is $262\pm5$~d. The epoch for peak
flux is MJD~50786. The predicted times of outbursts are indicated in
again Fig.~\ref{figasm}a. Only the last predicted outburst does not appear
to have materialized. The two WFC detections synchronize with the 2nd
and 5th outburst. A folded light curve (Fig.~\ref{figasm}c) shows an
average outburst profile which lasts $\sim$15~days. The average peak flux
is 4~mCrab.

\section{BATSE observations}
\label{secbatse}

\begin{figure}[t]
\psfig{figure=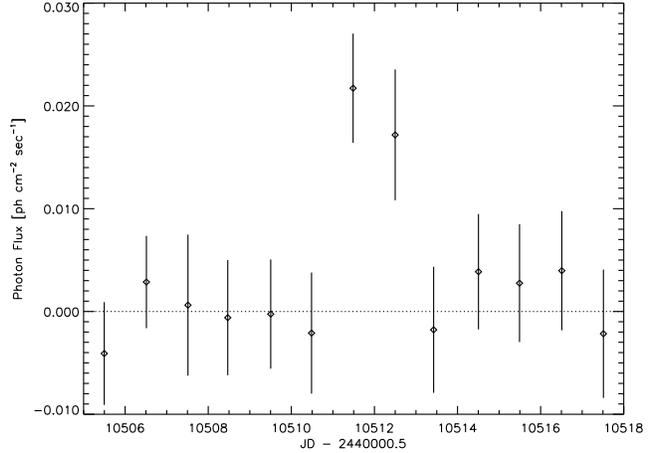,angle=90,width=\columnwidth}

\caption[]{BATSE light curve of the outburst of SAX J2239.3+6116 in 
March, 1997, for 20-100 keV with a one day time resolution.
\label{figbatse}}
\end{figure}

\begin{figure}[t]
\psfig{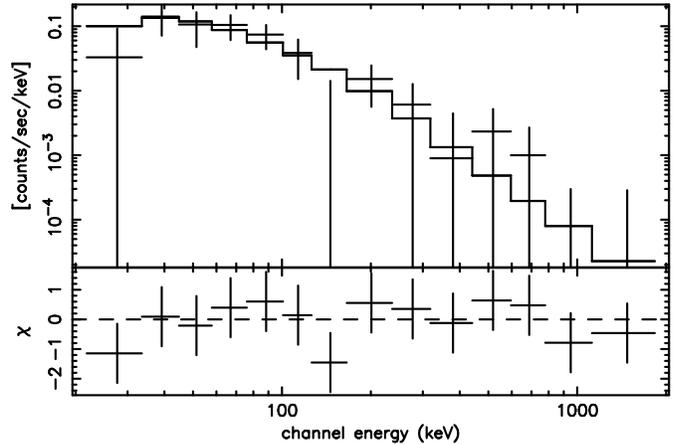}

\caption[]{Top panel: BATSE-measured spectrum of \bron (crosses) and
the fitted power-law model (histogram). Bottom panel: residuals between data
and model in units of sigma.
\label{figbatsespectrum}}
\end{figure}

Due to the hardness of the spectrum measured with the WFC 
a search of the BATSE data was made to see if \bron\ was detectable in the 
hard x-rays. The BATSE experiment onboard the Compton Gamma-Ray
Observatory (Fishman et al. 1989) using the
Large Area Detectors (LADs) can monitor the whole sky almost continuously in 
the energy range of 20 keV to 2 MeV with a typical daily 3$\sigma$ sensitivity
of better than 100~mCrab. Detector counting rates with 16 energy channel
energy resolution and a timing resolution of 2.048 seconds (CONT data) are
used for our data analysis.

To produce the \bron\ light curve, single step occultation data were taken
using a standard Earth occultation analysis technique used for monitoring hard
X-ray sources (Harmon et al. 1992). Interference from known bright sources
was removed. The single occultation step data were then fit with a power law
with a -2.0 photon index to determine daily flux measurements in the 20-100 keV
band. A time period of 30 days centered on each outburst was examined for
emission. Various time averages of the data were performed to search for a
possible signal in the data.

During the time of the March 1997 outburst we find evidence of a weak two day
outburst in the BATSE data (see Fig.~\ref{figbatse}). The 20-100 flux for
highest daily average corresponds to 0.080 times that of the Crab in the same
bandpass. It appears that there is a $\sim$3 day delay between the peak in
the BATSE data and that detected in the WFC. For any of the other outbursts
predicted throughout the BATSE operational period we do
not find evidence for hard x-ray emission, but we can not rule out the
presence of a source with a flux 0.050 times that of the Crab or less being
present in the data.

We have extracted spectra from the BATSE data for the March 1997 outburst.
During this time period the source was measured in three detectors. Spectra
and responses for each detector were created and a joint fit was made with a
simple power law. The fit was satisfactory, with a reduced chi-squared = 0.90 
(40 dof). The photon index is $-2.2\pm0.3$. In
Fig.~\ref{figbatsespectrum} is plot the spectrum in which the three data sets
have been grouped for plotting purposes. The 20-100 keV flux is
$1.4\times10^{-9}$~\ecs\ which is 0.080 times that of
the Crab in the same bandpass. Very little flux is detected above 100 keV.

\section{Optical observations}
\label{secopt}

\begin{figure}[t]
\psfig{figure=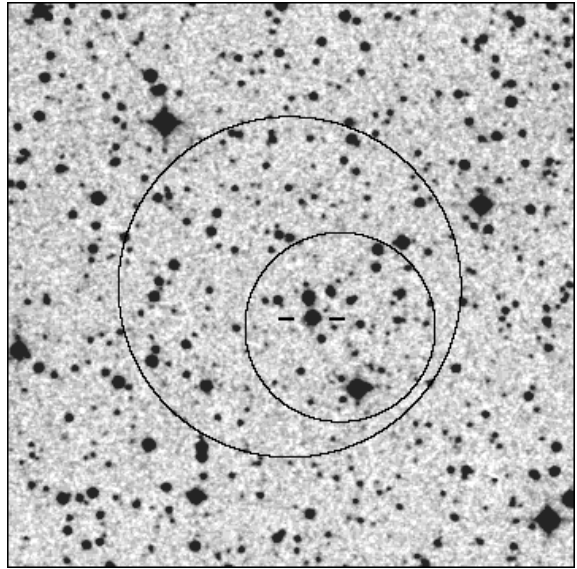,width=\columnwidth,clip=t}

\caption[]{Finding chart for the optical counterpart of SAX~J2239.3+6116
from the
Digitized Palomar Observatory Sky Survey.  The field is
$6^{\prime} \times 6^{\prime}$.  North is up, and east
is to the left.  The large and small circles are
the WFC and NFI error circles, respectively.
The optical counterpart is indicated by tick marks.  Its
position is $\alpha_{2000.0}$~=~$22^{\rm h}39^{\rm m}20^{\rm s}\!.90$,
$\delta_{2000.0}$~=~$+61^{\circ}16^{\prime}26^{\prime\prime}\!.8$,
\label{figopmap}}
\end{figure}

\begin{figure*}[t]
\psfig{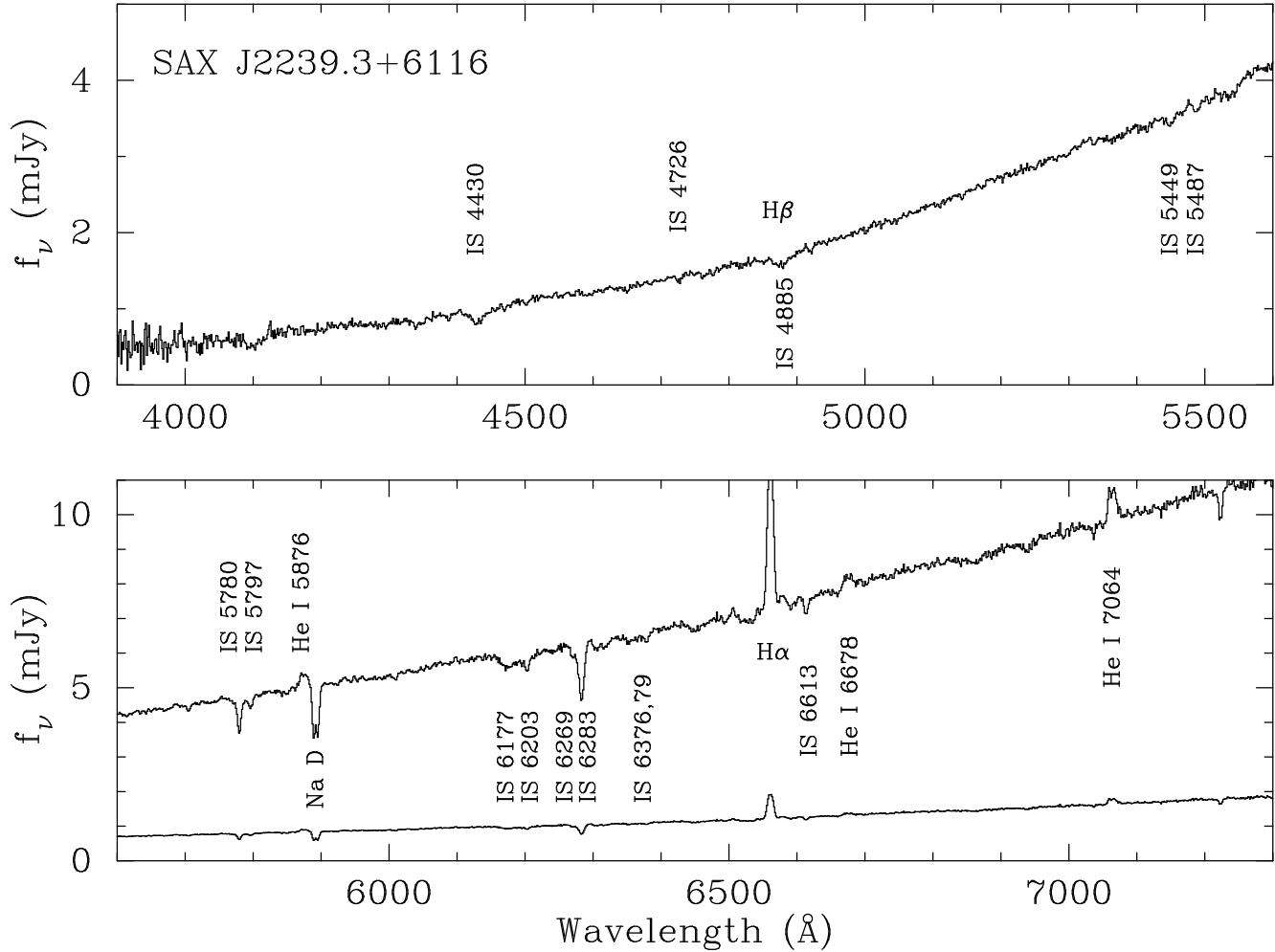}

\caption[]{Optical spectrum of SAX~J2239.3+6116
obtained on the KPNO 2.1m telescope on 1999
December 2 and 3.  In the lower panel, the flux scale refers to the upper
trace of the spectrum.  The lower trace is the same spectrum divided by a
factor of 6.
\label{figopspe}}
\end{figure*}

We looked for the optical counterpart of
SAX~J2239.3+6116 by obtaining spectra of the
three brightest stars in the NFI error circle on 1999
Dec. 2 and 3 UT, using the KPNO 2.1~m telescope and
Goldcam spectrograph. The date is 50~d before a predicted
outburst (the last indicated in Fig.~\ref{figasm}a).
The spectra covered the wavelength range
3900--7500~\AA\ at 5~\AA\ resolution. The 
second brightest star within the error circle
(Fig.~\ref{figopmap}) is the only one that has broad
emission lines and other spectral features that are
characteristic of a reddened Be star.  Be stars are often hard
X-ray sources because of emission from an unseen compact
companion such as a neutron star. On this basis we identify the Be
star with \bron.  This star is listed in the USNO--A2.0
catalog (Monet et al. 1996, Monet 1998), at a position
$\alpha_{2000.0} = 22^{\rm h}39^{\rm m}20^{\rm s}\!.90$,
$\delta_{2000.0} = +61^{\circ}16^{\prime}26^{\prime\prime}\!.8$ (uncertainty
$0^{\prime\prime}\!.3$),
which is $0.\!^{\prime}3$ from the NFI centroid and well within the
error circle.
The USNO--A2.0 catalog gives approximate magnitudes of
$B = 16.2$ and $R = 14.1$. 

The summed spectrum from 3000~s of exposure 
is shown in Fig.~\ref{figopspe}. We measure approximate
magnitudes $B = 16.5$, $V = 15.1$, and $R = 14.1$
from the spectrum. The estimated uncertainty in these numbers
is 0.2~mag.  The magnitudes are consistent with the USNO magnitudes.
The H$\alpha$ emission line has an equivalent width
(EW) of 6.7~\AA, a flux of $3.4 \times 10^{-14}$
ergs~cm$^{-2}$~s$^{-1}$, and a full-width at half maximum (FWHM) of
400~km~s$^{-1}$ after correcting for the instrumental resolution.
Emission lines of He~I~$\lambda$5876,
He~I~$\lambda$6678, and  He~I~$\lambda$7064 appear to be double-peaked,
with the peaks separated by 300~km~s$^{-1}$.
Double-peaked emission lines are commonly seen in Be star spectra,
where they are attributed to a circumstellar disk.

The reddening to the star may be estimated from many diffuse interstellar 
band (DIBs) that are apparent
in the spectrum (Fig.~\ref{figopspe}). We measure equivalent widths of
EW = 3.4~\AA, 3.2~\AA, and 2.3~\AA\ for $\lambda$4430,
Na~I D, and the $\lambda\lambda$5780,5797 blend, respectively.
From the calibrations of Herbig (1975) and T\"ug \& Schmidt-Kaler (1981)
we estimate $E(B-V)=1.4\pm0.2$. This is less than the value for the
{\em total\/} Galactic extinction as derived from the
hydrogen column density measurements through HI maps (Dickey \& Lockman 1995).
The integrated
H~I column density is $N_{\rm HI} = 1.0 \times 10^{22}$~cm$^{-2}$
in this direction. From the standard conversions
$N_{\rm H}/A_{\rm V} = 1.79 \times 10^{21}$~cm$^{-2}$~mag$^{-1}$
(Predehl \& Schmitt 1995) and $A_{\rm V} = 3.13 E(B-V)$, one obtains
$E(B-V)\simeq1.8$. Also, the reddening is less than determined from 
{\it IRAS\/} $100\, \mu$m dust maps (Schlegel et al. 1998) which
give $E(B-V)\simeq2.0$.

Assuming that $A_V = 3.13\,E(B-V)$,
the dereddened magnitudes from the spectrum become
$B = 10.7$, $V = 10.7$ and $R = 10.9$. Since these are somewhat
redder than the colors of an early B star, we suspect that there
may be some additional circumstellar extinction, and that 
dereddened $V\approx9.7$ is a realistic estimate.  Also, if we assume an
average $A_V = 1$ mag~kpc$^{-1}$, then the distance to
SAX~J2239.3+6116 can be estimated as 4.4~kpc.  The absolute
visual magnitude would then be --3.5, as expected for a star
in the range B0~V to B2~III.

In a 16 square-degree area around the NFI position, there are
3 cataloged Be stars brighter than $V=10$. None of these are
coincident with the NFI error box.

\section{Association with other sources of high-energy emission}
\label{secoth}

\begin{figure}[t]
\psfig{figure=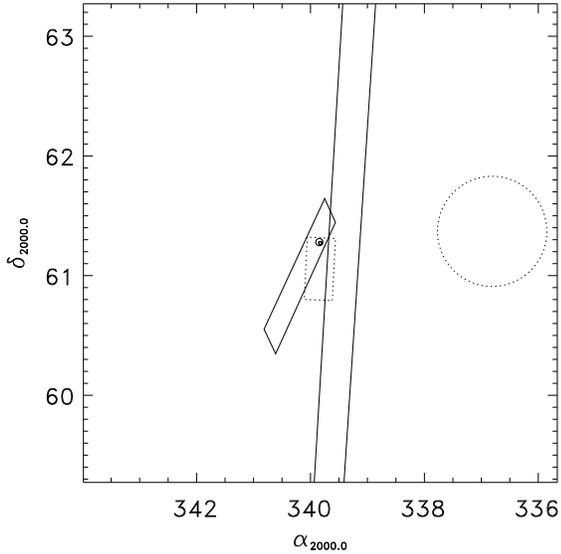,width=\columnwidth,clip=t}

\caption[]{Map of relevant high-energy sources. The large error circle
to the right is
that of \egret, the two straight lines refer to the fast transient AT~2238+584,
the dashed quadrilateral to 3A~2237+608, the solid quadrilateral to 
4U~2238+60, the two
small circles to the WFC error box (large) and NFI error box (small) of
\bron.
\label{figmap}}
\end{figure}

In Fig.~\ref{figmap}, the WFC- and MECS-determined positions of \bron\ are
shown, as well as those of three other possibly related X-ray sources. It 
appears likely that \bron\ is the same source as the one seen in
the early seventies with {\rm Uhuru\/} (4U~2238+60, Forman et al. 1978),
OSO-7 (1M~2233+595, Markert et al. 1979), and {\em Ariel-V\/} (3A~2237+608, 
Warwick et al. 1981). 4U~2238+60 was detected at a level of 
$(7\pm1)\times10^{-11}$~\ecs\ (2-10 keV, statistical error only) which is about
4 times lower than the peak of the March 1997 outburst. This is not really
surprising
because neither measurement encompassed the whole outburst. The 
{\em Ariel\/} source (Warwick et al. 1981) is classified as an irregular 
variable source during the 1974-1980 operational period
with a measured peak flux of $2\times10^{-10}$~\ecs\ (2-10 keV)
which is also of 
the same order of magnitude as \bron\ and 4U~2238+60. 

There are two other interesting sources which have error boxes that are 
close to that of \bron\ but are formally inconsistent. The first one
is the fast transient AT~2238+584 which was seen in July 1975 with the 
sky-survey instrument on {\em Ariel V} (Pye \& McHardy 1983). The outburst 
was not covered completely. The maximum measured flux was
$(1.6\pm0.2)\times10^{-9}$~\ecs\ (2-10 keV) which is 5 times as bright as
for \bron. The e-folding decay time of 
AT~2238+584 was 0.78~d. The other inconsistent source is \egret\ 
which is one of the unidentified EGRET sources (Hartman et al. 1999). We 
mention this because \bron\ is the bright X-ray source closest
to \egret\ and because \bron\ has quite a hard spectrum in
the X-ray range. Nevertheless, the distance between the centroids of
both sources is 87\arcmin\ while the 95\% confidence error region of
\egret\ is three times smaller (28\arcmin).

\section{Discussion}
\label{secdiscuss}

\bron\ is an X-ray transient which often recurs with a periodicity
of 262~d. Because of the Be-star nature of the likely optical counterpart, the
periodicity may be identified with the orbital period of the binary.
In such a Be X-ray binary system, the transient nature of the X-rays is
thought to arise from a combination of episodic mass loss by
the Be star, and an eccentric binary orbit. The compact object orbiting
the Be star will show enhanced X-ray emission near periastron where
it accretes more matter due to an enhanced wind density (e.g., review
by White et al. 1995). The 2-10 keV flux
history shows fluctuations of a factor of 10$^3$. Possibly the true
dynamic range is larger. The likely counterpart at an estimated distance
of 4.4~kpc implies that the 2-10 keV luminosity ranged between
1$\times10^{33}$~\lum\ and $7\times10^{35}$~\lum. It seems plausible that
\bron\ was more or less active over the last decades. The peak flux is so
low that it can easily be missed. The spectral type of the counterpart, the
resulting luminosity and the transient behavior is consistent with the
identification of \bron\ as a Be X-ray binary. This is also confirmed by
the value for the X-ray to optical color index
$\xi=B + 2.5{\rm log}(F_X(\mu{\rm Jy}))$
(as defined by Van Paradijs \& McClintock
1995) where $B$ is the dereddened $B$ magnitude and $F_X$ the 2-10 keV X-ray
flux. For $B=10.7$ and $F_X=0.03~\mu$Jy (during quiescence) $\xi=6.9$. If
$B$ would be 9.7 (to correct for additional circumstellar extinction) then
$\xi=5.9$. These are perfectly normal values for Be X-ray binaries in
quiescence and three times lower than for low-mass X-ray binaries.

Usually Be X-ray binaries contain
an X-ray pulsar (e.g., review by White et al. 1995). \bron\ does not appear
to contain one with a period shorter than 10$^3$~sec, with a fairly sensitive
upper limit of 10\% on the amplitude. This may be explained by the fact that 
the pulse period is longer than 10$^3$~sec. Corbet (1984) discovered a strong
correlation between orbital and pulse period for Be X-ray binaries. If \bron\
adheres to this correlation, the orbital period of 262~d implies a
pulse period near to 10$^3$~sec. More dedicated sensitive X-ray observations
during future outbursts are obviously needed to study this system in more
detail.

\begin{acknowledgements}
We are grateful to the staff of the {\em BeppoSAX\/} Science Data
Center for processing the vast amounts of raw data from WFC secondary mode
observations which were the basis for the study presented in this paper,
and thank Lucien Kuiper for help in analyzing the MECS data with the
maximum likelihood method. {\em BeppoSAX\/} is a joint Italian and Dutch
program.

\end{acknowledgements}

\end{document}